\documentclass{IEEEtran4PSCC}
\ifCLASSINFOpdf
   \usepackage[pdftex]{graphicx}
\else
   \usepackage[dvips]{graphicx}
\fi
\usepackage[cmex10]{amsmath}

\usepackage{amsmath, amssymb, amsfonts, amsthm, booktabs, cancel, comment, enumitem, eurosym, float, graphicx, mathtools, multirow, nicefrac, placeins, soul, steinmetz, stfloats, siunitx, svg, url, xcolor, tabularx}%
\usepackage{commath}
\usepackage{cite}
\usepackage{subfigure}
\usepackage{hyperref}
\usepackage{nomencl}
\usepackage{cleveref}
\makenomenclature

\floatstyle{ruled}
\newfloat{model}{thp}{lop}
\floatname{model}{\textsc{Model}}

\makeatletter
\renewcommand*{\fnum@model}{\fname@model}
\makeatother

\makeatletter
\let\old@ps@headings\ps@headings
\let\old@ps@IEEEtitlepagestyle\ps@IEEEtitlepagestyle
\def\psccfooter#1{%
    \def\ps@headings{%
        \old@ps@headings%
        \def\@oddfoot{\strut\hfill#1\hfill\strut}%
        \def\@evenfoot{\strut\hfill#1\hfill\strut}%
    }%
    \def\ps@IEEEtitlepagestyle{%
        \old@ps@IEEEtitlepagestyle%
        \def\@oddfoot{\strut\hfill#1\hfill\strut}%
        \def\@evenfoot{\strut\hfill#1\hfill\strut}%
    }%
    \ps@headings%
}
\makeatother

\psccfooter{%
        \parbox{\textwidth}{\hrulefill \\ \small{23rd Power Systems Computation Conference} \hfill \begin{minipage}{0.2\textwidth}\centering \vspace*{4pt} \includegraphics[scale=0.06]{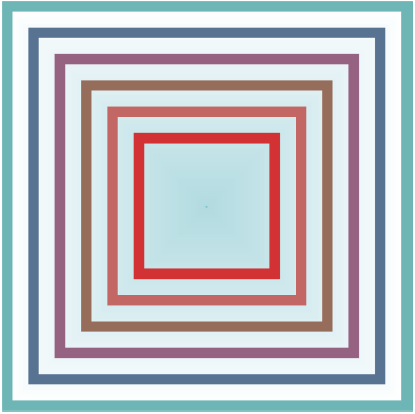}\\\small{PSCC 2024} \end{minipage} \hfill \small{Paris, France --- June 4 -- 7, 2024}}%
}

\begin{document}
%


\title{Interpreting the Value of Flexibility in AC Security-Constrained Transmission Expansion Planning via a Cooperative Game Framework}

\author{\IEEEauthorblockN{
Andrey Churkin\IEEEauthorrefmark{1},
Wangwei Kong\IEEEauthorrefmark{2}, 
Mohammad~Iman~Alizadeh\IEEEauthorrefmark{3},
Florin~Capitanescu\IEEEauthorrefmark{3},\\
Pierluigi Mancarella\IEEEauthorrefmark{1}\textsuperscript{,}\IEEEauthorrefmark{4} and
Eduardo~A.~Martínez~Ceseña\IEEEauthorrefmark{1}\textsuperscript{,}\IEEEauthorrefmark{5}}
\vspace{2\jot}
\IEEEauthorblockA{\IEEEauthorrefmark{1}Department of Electrical and Electronic Engineering, The University of Manchester, U.K.}
\IEEEauthorblockA{\IEEEauthorrefmark{2}National Grid Electricity Transmission, U.K.}
\IEEEauthorblockA{\IEEEauthorrefmark{3}Luxembourg Institute of Science and Technology (LIST), Luxembourg}
\IEEEauthorblockA{\IEEEauthorrefmark{4}Department of Electrical and Electronic Engineering, The University of Melbourne, Australia}
\IEEEauthorblockA{\IEEEauthorrefmark{5}Tyndall Centre for Climate Change Research, U.K.}
}

\maketitle

\begin{abstract}

Security-constrained transmission expansion planning (SCTEP) is an inherently complex problem that requires simultaneously solving multiple contingency states of the system (usually corresponding to \mbox{N-1} security criterion). Existing studies focus on effectively finding optimal solutions; however, single optimal solutions are not sufficient to interpret the value of flexibility (e.g., from energy storage systems) and support system planners in well-informed decision making. 
In view of planning uncertainties, it is necessary to estimate the contributions of flexibility to various objectives and prioritise the most effective investments.
In this regard, this work introduces a SCTEP tool that enables interpreting the value of flexibility in terms of contributions to avoided load curtailment and total expected system cost reduction.
Inspired by cooperative game theory, the tool ranks the contributions of flexibility providers and compares them against traditional line reinforcements. This information can be used by system planners to prioritise investments with higher contributions and synergistic capabilities.

\end{abstract}

\begin{IEEEkeywords}
Flexibility, security-constrained optimal power flow, Shapley value, stochastic optimisation, transmission network planning.
\vspace{-2\jot}
\end{IEEEkeywords}

\thanksto{\noindent Submitted to the 23rd Power Systems Computation Conference (PSCC 2024).
Corresponding author: Andrey Churkin \url{https://andreychurkin.ru/}}

\section{Introduction: The Need for Interpretable Security-constrained Planning
}\label{Section: introduction}

Security-constrained transmission expansion planning (SCTEP) is becoming increasingly important as the integration of uncertain renewable energy sources (RES) accelerates \cite{GOMES2019411}. To guarantee cost-effective and reliable (e.g., N-1 secure) operation of future power systems, it is necessary to develop accurate SCTEP models capable of identifying potential congestion and voltage issues and finding optimal investments in assets (such as line reinforcement) and flexibility (energy storage systems, demand response programs, etc.).

Power system planning with N-1 security constraints is an inherently complex problem, as it requires simultaneously solving normal system conditions alongside multiple contingency states. This is recognised by existing security-constrained optimal power flow (SCOPF) literature, as SCOPF models used for verifying the feasibility of network plans have high computational costs associated with contingencies, large numbers of variables and constraints, nonlinearities and nonconvexities of the optimisation problems \cite{CAPITANESCU2011,Florin_in_book2,ALIZADEH2022Envisioning}.

To overcome the above challenges, existing studies focus on simplifying the SCTEP formulation, e.g., via power flow linearisations and metaheuristic algorithms \cite{Zhang20212,MORQUECHO2023,Iman2022_tractable}.
These models allow system planners to approximate single optimal SCTEP solutions and develop investment portfolios.
However, a single optimal solution is not sufficient to interpret the value of flexibility and support well-informed decision making. That is, it is unclear how flexibility providers contribute to the economic efficiency and N-1 security of a system, and how effective they are compared to traditional line reinforcement. Without this information, system planners cannot prioritise investment options (to deploy first the solutions that bring more value) and analyse their potential synergies. Prioritising investments becomes especially important in multi-stage planning, where uncertainties from later stages or years can compromise the adequacy of the planning decisions being made now.

In this regard, this work introduces an open-source SCTEP tool for interpreting the value of flexibility:
\vspace{-1.5\jot}
\begin{center}
{\fontfamily{qcr}\selectfont
\url{https://github.com/AndreyChurkin/iSCTEP/}}
\end{center}
\vspace{-1.5\jot}
The tool enables interpreting the value of flexibility in terms of contributions to avoided load curtailment and total expected system cost reduction. 
Inspired by cooperative game theory \cite{Churkin2021review}, the tool ranks the contributions of flexibility providers and compares them against traditional line reinforcements.
This information allows system planners to prioritise investments with higher contributions and synergistic capabilities, i.e., the largest contributions in combination with other investments.

Specifically, the tool formulates a cooperative game among selected investment options (line reinforcements and flexibility providers) and iteratively solves a SCTEP model to estimate the value of investments in different coalitions (combinations of investments).
The planning model is based on an exact nonlinear stochastic AC SCOPF formulation, adapted from previous work \cite{ALIZADEH2022Envisioning}, which enables finding accurate AC-feasible solutions for multiple post-contingency states and incorporating uncertainties from RES.
Then, marginal contributions of investments to coalitions are estimated and the Shapley value, a popular valuation concept from cooperative game theory \cite{Churkin2021review}, is computed to provide a single-valued estimation of the usefulness of investments.
As demonstrated through simulations in Section~\ref{Section: results}, cooperative game theory provides a natural framework for analysis of investment options in the SCTEP problem, giving unique insights into the contributions of investments and their synergistic capability.

Similar coalitional analyses have been done in power systems research.
In \cite{Hasan2018}, solution concepts from cooperative game theory were applied for the identification of critical parameters affecting power system small-disturbance stability.
In \cite{CAO2022}, the Shapley value was used to estimate the impact of various components of flexible multi-energy systems on their reliability. 
In \cite{Churkin2023tracing}, the usefulness of the Shapley value was demonstrated for estimating the criticality of flexibility providers in active distribution networks.
The proposed tool borrows the ranking and tracing ideas from \cite{Hasan2018,CAO2022,Churkin2023tracing}, but further develops them to address the flexibility valuation problem in transmission network planning with security constraints.
In the context of transmission expansion planning, cooperative game theory has been applied to analyse the benefits of transmission expansion projects \cite{BANEZCHICHARRO2017beneficiaries,BANEZCHICHARRO2017benefits} and to rank the usefulness of fast-ramping flexibility providers \cite{Kristiansen2018}.
Even though exploring similar concepts, studies \cite{BANEZCHICHARRO2017beneficiaries,BANEZCHICHARRO2017benefits,Kristiansen2018} are different from current work as they: (i) rely on linear optimisation models, (ii) do not consider contingencies, and (iii) do not compare investment in flexibility against line reinforcement options.

The usefulness of the Shapley value has also been recognised beyond power systems research. For example, it has been demonstrated that the Shapley value and its approximations can be used to effectively interpret machine learning models and solve data valuation problems \cite{Jia2020,Mitchell2022}.\footnote{The scalability of the Shapley value is often considered its main limitation as the number of possible coalitions in a cooperative game, $2^N$, increases exponentially with the number of players $N$. Nevertheless, several approaches have been developed to overcome this issue. An overview of such methods can be found in \cite{Churkin2023tracing}, section IV ``Scalability and Applicability Issues".}
Similar to the SCTEP problem, complex machine learning models, typically formulated as black-box models, can provide a single solution to a classification or prediction problem. However, a single solution is not enough to interpret the outcome of the learning model and the value of the input data and its features. Therefore, interpretable machine learning models based on the Shapley value are getting increasing attention.


This work combines recent advances in SCOPF and SCTEP modelling and concepts from interpretable models (using cooperative game theory) and makes the following contributions:
\begin{itemize}
    \item A new SCTEP planning tool is developed for interpreting the value of flexibility in terms of contributions to avoided load curtailment and total expected system cost reduction via the coalitional analysis of investment options, allowing system planners to prioritise investments with higher contributions and synergistic capabilities.
    To the authors' knowledge, no such analysis has been performed in the context of SCTEP. 
    Compared to relevant studies \cite{BANEZCHICHARRO2017beneficiaries,BANEZCHICHARRO2017benefits,Kristiansen2018}, the developed tool is based on the exact nonlinear stochastic AC SCOPF formulation, adapted from \cite{ALIZADEH2022Envisioning}.
    \item The proposed cooperative game formulation considers investment options as players, which enables comparing the value of flexibility against traditional line reinforcement.
    In contrast to works such as \cite{Hasan2018,CAO2022,Churkin2023tracing}, the tool not only identifies critical system components and contingencies but also suggests investments that effectively deal with potential system operation problems.
\end{itemize}

The proposed SCTEP tool is demonstrated with a simple 5-bus system and a realistic 30-bus UK electricity transmission system. The performed coalitional analysis for the selected investment options provides the following findings:
\begin{itemize}
\item Investments in flexibility and line reinforcement offer vastly different values depending on the objectives of the system planner. Flexibility providers consistently reduce load curtailment in many coalitions, as they can directly reduce peak load demand at specific locations. Line reinforcements tend to offer lower costs by facilitating the transfer of power from cheaper generators.
\item Individual contributions of investment options cannot always be used to correctly estimate the entire range of possible contributions. That is, traditional sensitivity analyses (exclusion of certain investments) can lead to an incorrect assessment of the investment's value and synergistic capability. A thorough coalitional analysis is required to prioritise investment options.
\end{itemize}


\section{Modelling Framework: Coalitional Analysis for the SCTEP Problem}\label{Section: models}
This section introduces the SCTEP problem formulation as a nonlinear optimisation model. Then, a cooperative game among investments is formulated to interpret their value and identify investments with the highest contributions to the selected objectives. Finally, an overview of the proposed modelling framework is presented to explain the interactions between the SCTEP model and the cooperative game.

\subsection{SCTEP Formulation}
A nonlinear stochastic AC SCOPF model, originally introduced and tested in \cite{ALIZADEH2022Envisioning}, is adapted in this work to develop the SCTEP tool. The SCOPF model is extended by explicitly including planning decision variables: additional line capacities due to reinforcements and capacities of flexibility providers.\footnote{Note that investment options are defined as continuous variables (MVA capacity increase for lines and flexibility providers), while in practice these variables are discrete. The inclusion of integer and binary variables will be considered in future research.
}
The complete model formulation is presented in \eqref{ACSCOPF: p_nm}-\eqref{ACSCOPF: curt limit}, where variables and constraints are defined for each scenario $s$ and state of the system $k$.
To ease the notation (to avoid duplication of similar constraints), all considered system states are denoted by $k$. Yet, within the tool, $k=0$ indicates the normal operation and $k\geq1$ indicates contingency states.\footnote{In this work, only line contingencies corresponding to the N-1 security criterion are explicitly included in the formulation. However, the model can be extended by specifying additional contingency states due to the failure of generators and other equipment.}
\begin{model*}[t]
\caption{Stochastic AC security-constrained power flow for SCTEP \hfill [continuous NLP, QCP]}
\label{ACSCOPF}
\begin{subequations} 
\vspace{-2\jot}
\begin{IEEEeqnarray}{llll}
\textbf{Variables:} \text{ (for scenario $s \in \mathcal{S}$ and state $k \in \mathcal{K}$)}\IEEEnonumber\\
e^k_{b,s}, f^k_{b,s} \hspace{30.7mm} \text{real and imaginary parts of complex voltage} &\forall b \in \mathcal{B} \IEEEnonumber\\
p_{bm,s}^k, q_{bm,s}^k \hspace{25.7mm}  \text{active and reactive power flows} &\forall (b,m) \in \mathcal{L} \IEEEnonumber\\
P_{b,g,s}^k, Q_{b,g,s}^k \hspace{24.5mm} \text{active and reactive power production of generators} &\forall b \in \mathcal{B}, g \in \mathcal{G}\IEEEnonumber\\
P^{\uparrow,k}_{b,f,s}, P^{\downarrow,k}_{b,f,s}, Q^{\uparrow,k}_{b,f,s}, Q^{\downarrow,k}_{b,f,s} \hspace{4.4mm} \text{increase and decrease of active/reactive power by flexibility providers} \quad &\forall b \in \mathcal{B}, f \in \mathcal{F} \IEEEnonumber\\
{LC}^k_{b,s}, {RC}^k_{b,s} \hspace{23.5mm} \text{active load curtailment and RES power curtailment} &\forall b \in \mathcal{B} \IEEEnonumber\\
{LI}_{bm} \hspace{34.8mm} \text{additional line capacity due to reinforcement} &\forall (b,m) \in \mathcal{L} \IEEEnonumber\\
{FI}_{b,f} \hspace{34.5mm} \text{additional capacity of flexibility providers due to investments} &\forall b \in \mathcal{B}, f \in \mathcal{F} \vspace{2\jot} \IEEEnonumber\\
\textbf{Constraints:} \text{ (for scenario $s \in \mathcal{S}$ and state $k \in \mathcal{K}$)}\IEEEnonumber\\
p_{bm,s}^k = ({e^k_{b,s}}^2+{f^k_{b,s}}^2)G_{bm} - (e^k_{b,s} e^k_{m,s} + f^k_{b,s} f^k_{m,s})G_{bm} - (f^k_{b,s} e^k_{m,s} - e^k_{b,s} f^k_{m,s})B_{bm} \qquad\quad &\forall (b,m) \in \mathcal{L} \quad \label{ACSCOPF: p_nm}\\
q_{bm,s}^k = -({{e^k_{b,s}}}^2+{f^k_{b,s}}^2)B_{bm} + (e^k_{b,s} e^k_{m,s} + f^k_{b,s} f^k_{m,s})B_{bm} - (f^k_{b,s} e^k_{m,s} - e^k_{b,s} f^k_{m,s})G_{bm} \quad\quad &\forall (b,m) \in \mathcal{L} \qquad \label{ACSCOPF: q_nm}\\
\smashoperator{\sum_{g \in \mathcal{G}}}P_{b,g,s}^k + P^R_{b,s} - {RC}^k_{b,s} - P^D_{b,s} + {LC}^k_{b,s} + \smashoperator{\sum_{f \in \mathcal{F}}}(P^{\uparrow,k}_{b,f,s} - P^{\downarrow,k}_{b,f,s})  -\smashoperator{\sum_{(b,m) \in \mathcal{L}}}p_{bm,s}^k = 0 &\forall b \in \mathcal{B}\quad \label{ACSCOPF: balance_p}\\
\smashoperator{\sum_{g \in \mathcal{G}}}Q_{b,g,s}^k - Q^D_{b,s} + \smashoperator{\sum_{f \in \mathcal{F}}}(Q^{\uparrow,k}_{b,f,s} - Q^{\downarrow,k}_{b,f,s}) -\smashoperator{\sum_{(b,m) \in \mathcal{L}}}q_{bm,s}^k = 0 &\forall b \in \mathcal{B}\quad \label{ACSCOPF: balance_q}\\
\underline{P}_{b,g} \leq P_{b,g,s}^k \leq \overline{P}_{b,g}, \quad \underline{Q}_{b,g} \leq Q_{b,g,s}^k \leq \overline{Q}_{b,g} &\forall b \in \mathcal{B}, g \in \mathcal{G} \label{ACSCOPF: p_lim}\\
{p_{bm,s}^k}^2 + {q_{bm,s}^k}^2 \leq {(\overline{\mathcal{S}}_{bm} + {LI}_{bm})}^2 &\forall (b,m) \in \mathcal{L} \label{ACSCOPF: Smax}\\
{\underline{V}_n}^2 \leq {e^k_{b,s}}^2+{f^k_{b,s}}^2 \leq {\overline{V}_n}^2 &\forall k \in \mathcal{K} \label{ACSCOPF: vmax}\\
0 \leq P^{\uparrow,k}_{b,f,s} \leq \overline{P}^{\uparrow}_{b,f} + {FI}_{b,f}, \quad 0 \leq P^{\downarrow,k}_{b,f,s} \leq \overline{P}^{\downarrow}_{b,f} + {FI}_{b,f} &\forall b \in \mathcal{B}, f \in \mathcal{F} \label{ACSCOPF: p_up}\\
0 \leq Q^{\uparrow,k}_{b,f,s} \leq \overline{Q}^{\uparrow}_{b,f} + {FI}_{b,f}, \quad 0 \leq Q^{\downarrow,k}_{b,f,s} \leq \overline{Q}^{\downarrow}_{b,f} + {FI}_{b,f} &\forall b \in \mathcal{B}, f \in \mathcal{F} \label{ACSCOPF: q_up}\\
0 \leq LI_{bm} \leq \overline{LI}_{bm} &\forall (b,m) \in \mathcal{L} \label{ACSCOPF: line investments limit}\\
0 \leq FI_{b,f} \leq \overline{FI}_{b,f} &\forall b \in \mathcal{B}, f \in \mathcal{F} \label{ACSCOPF: flex investments limit}\\
0 \leq {LC}^k_{b,s} \leq P^D_{b,s}, \quad 0 \leq {RC}^k_{b,s} \leq P^R_{b,s} &\forall b \in \mathcal{B} \label{ACSCOPF: curt limit}
\end{IEEEeqnarray}
\end{subequations}
\end{model*}

Active and reactive power flows are defined in \eqref{ACSCOPF: p_nm} and \eqref{ACSCOPF: q_nm}, where $G_{bm}$ and $B_{bm}$ are the conductance and susceptance of lines, respectively.
Constraints \eqref{ACSCOPF: balance_p} and \eqref{ACSCOPF: balance_q} represent active and reactive power balance
equations for each bus $b \in \mathcal{B}$ of the system. 
In \eqref{ACSCOPF: p_lim}, limits of active and reactive power are defined for each generator $g$.
Network operation constraints, congestion and voltages, are imposed in \eqref{ACSCOPF: Smax} and \eqref{ACSCOPF: vmax}.
The initial apparent power flow limits for lines are denoted by $\overline{\mathcal{S}}_{bm}$ corresponding to the line capacity before reinforcement.
Operation limits of flexibility providers are defined by the set of constraints \eqref{ACSCOPF: p_up}, \eqref{ACSCOPF: q_up}.
Investments in line reinforcement and flexibility providers are capped with \eqref{ACSCOPF: line investments limit} and \eqref{ACSCOPF: flex investments limit}.
Finally, \eqref{ACSCOPF: curt limit} defines limits of load curtailment and RES power curtailment. Note that curtailments of RES and loads (and the associated penalties) are introduced to prevent infeasibility of the AC SCOPF model at the system operation stage.

Two objective functions are considered in this work for the SCTEP problem formulation. The first objective is the minimisation of load curtailments, ${LC}^k_{b,s}$, for all scenarios $s$ and states $k$:
\begin{IEEEeqnarray}{lll}
    \label{Objective: LC}
    \min \enskip \smashoperator{\sum_{s \in \mathcal{S}}} \ \smashoperator{\sum_{k \in \mathcal{K}}} \ \smashoperator{\sum_{b \in \mathcal{B}}} {LC}^k_{b,s}
\end{IEEEeqnarray} 

The second objective is the minimisation of the total expected system cost:
\begin{IEEEeqnarray}{lll}
    \label{Objective: total cost}
    \min \enskip \smashoperator{\sum_{s \in \mathcal{S}}} \smashoperator{\sum_{k \in \mathcal{K}}} \pi_{s}^k \Bigg[ 
     \smashoperator{\sum_{b \in \mathcal{B}}} \bigg[
    \smashoperator{\sum_{g \in \mathcal{G}}}c^{\text{gen}}_{b,g}(P^k_{b,g,s}) \IEEEnonumber\\
    \quad
    + {LC}^k_{b,s} c^\text{curt}_b
    + {RC}^k_{b,s} c^\text{curt}_{R} 
    + \smashoperator{\sum_{f \in \mathcal{F}}}(P^{\uparrow,k}_{b,f,s} + P^{\downarrow,k}_{b,f,s})c^{\text{flex}}_{b,f} \bigg]  
    \Bigg] \quad \IEEEnonumber\\
    \quad + \smashoperator{\sum_{(b,m) \in \mathcal{L}}}{LI}_{bm} c_{bm}^\text{inv}
    + \smashoperator{\sum_{b \in \mathcal{B}}}\smashoperator{\sum_{f \in \mathcal{F}}}{FI}_{b,f} c_f^\text{inv}
\end{IEEEeqnarray}


This objective comprises the cost of generation (given by function $c^{\text{gen}}$), penalties associated with load curtailment and RES power curtailment, $c^\text{curt}_b$, $c^\text{curt}_R$, cost of flexible power production, $c^{\text{flex}}_{b,f}$, and investments in line reinforcement and flexibility providers (with levelised costs $c_{bm}^\text{inv}$ and $c_f^\text{inv}$). The probability of occurrence of scenario $s$ for state $k$ is given by $\pi_{s}^k$. In the simulations presented in this work, for simplicity reasons, all scenarios and contingencies are assumed equiprobable. The probability of a contingency state is assumed to be 0.05 while the normal state has a probability of 0.95.
In real-world applications, a more accurate estimation of probabilities for each contingency might be required \cite{ContingencyProbability2006}.

Note that the introduced AC SCOPF model \eqref{ACSCOPF: p_nm}-\eqref{ACSCOPF: curt limit} with objective functions \eqref{Objective: LC}, \eqref{Objective: total cost} is an instance of nonlinear programming (NLP). Specifically, the planning problem is formulated as a quadratically constrained programming (QCP) problem, which is generally nonlinear and nonconvex. Yet, due to the omission of investment binary variables, the formulated problem is continuous. Therefore, it can be solved with software for nonlinear continuous systems, e.g., Ipopt solver.






\subsection{Coalitional Analysis: Interpreting SCTEP Solutions via Cooperative Game Theory}

Cooperative game theory offers a natural framework to analyse the contributions of investment options to the defined objectives \cite{Churkin2021review,BANEZCHICHARRO2017beneficiaries,BANEZCHICHARRO2017benefits,Kristiansen2018}. For this purpose, investments are represented as players jointly contributing to different coalitions (combinations of investments).
Specifically, a cooperative game $(N;v)$ in SCTEP can be formulated as follows:
\begin{itemize}
	\item $N$ is a finite set of players (investment options considered in the STEP problem). A subset of $N$ is called a coalition. The largest possible coalition containing all players is called the grand coalition. The collection of all coalitions is denoted by $2^N$.
	\item $v$ $:$ ${2^N} \rightarrow \mathbb{R}$ is the characteristic function associating each coalition $S$ with a real number $v(S)$, which is a metric describing the value of a coalition.
\end{itemize}

In this work, the defined objective functions \eqref{Objective: LC} and \eqref{Objective: total cost} are used to characterise the value of coalitions in terms of load curtailment and total expected system cost. That is, the optimal solution for the SCTEP problem with a coalition of investments $S$ defines the value of the coalition. 
The difference between the SCTEP problems for possible coalitions stems from available investments defined by the set of players in each coalition: if the investment is part of a coalition, it can be used to optimise the transmission plan for that coalition.

Then, the marginal contribution to coalition $S$ by player $i$ is estimated as the difference in the coalition's value with and without the player:
\begin{IEEEeqnarray}{lll}
    \label{CGT: MC_i}
    {MC(S)}_i = v(S \cup \{i\})-v(S) \quad \forall i \in S \quad \forall S \subseteq N 
\end{IEEEeqnarray} 

By comparing the differences between multiple optimal SCTEP solutions, marginal contributions calculated in \eqref{CGT: MC_i} represent maximum avoided load curtailment and total expected cost reduction driven by the investments.
Analysis of contributions of the investments to all ${2^N}$ coalitions enables accurately estimating their usefulness and synergistic capabilities, i.e., potential contributions in combination with other investments.
As a single-valued estimation of the contributions by player $i$, the Shapley value can be calculated using \eqref{CGT: Shapley}. The Shapley value provides the weighted average of players' marginal contributions to all possible coalitions, where $\lvert N \rvert$ denotes the total number of players in a cooperative game and $\lvert S \rvert$ is the number of players in coalition $S$.
\begin{IEEEeqnarray}{l}
    \label{CGT: Shapley}
    Sh_i=\sum_{S \subseteq N \setminus \{i\} } \frac{|S|!(|N|-|S|-1)!}{|N|!} {MC(S)}_i
    \enskip\enskip\enskip
\end{IEEEeqnarray}

A simplified analysis of coalitions can be done by considering only players' contributions to the grand coalition and their individual contributions (to coalitions with only one player).
That is, a truncated coalitional structure can be considered, for example, to perform the initial screening of investment options and select investments with the highest contributions for more detailed coalitional analysis.
This corresponds to the traditional planning approach, where the system planner can perform a sensitivity analysis of the optimal solution by excluding one of the investment options (or excluding all but one investment). 
However, as will be demonstrated by the simulations, such a simple sensitivity analysis can lead to an incorrect assessment of the investment's value and its synergistic capability.

\subsection{Modelling Framework Overview}
\begin{figure}
    \centering
    \includegraphics[width=0.95\columnwidth]{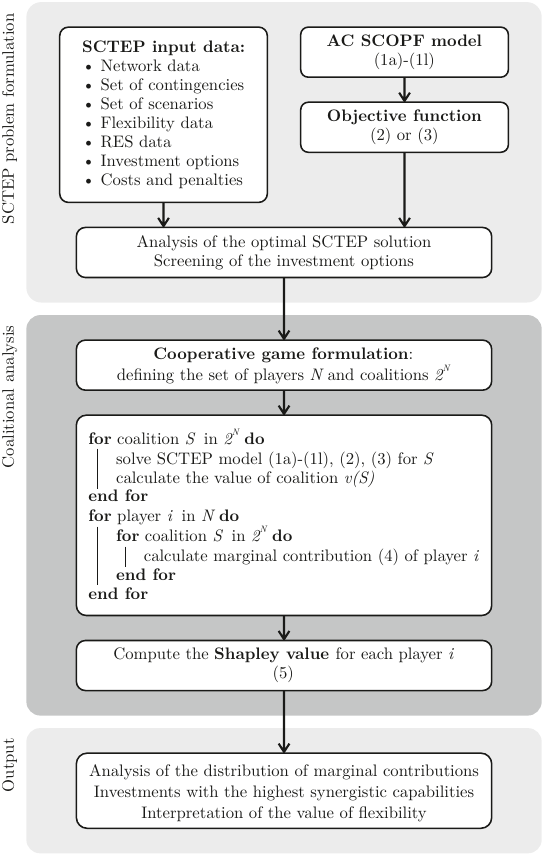}
    \caption{Framework diagram.}
    \label{Fig: flowchart}
\end{figure}

The proposed framework requires interactions between the SCTEP model \eqref{ACSCOPF: p_nm}-\eqref{ACSCOPF: curt limit}, \eqref{Objective: LC}, \eqref{Objective: total cost} and the cooperative game formulation $(N;v)$.
To elucidate these interactions, the framework diagram is presented in Figure~\ref{Fig: flowchart}. The diagram is divided into three blocks corresponding to (i) the SCTEP model formulation and data preparation, (ii) coalitional analysis based on the cooperative game formulation, and (iii) output and subsequent analysis.

In the first block, the case study data and the AC SCOPF model \eqref{ACSCOPF: p_nm}-\eqref{ACSCOPF: curt limit} are used to find the optimal solution for the SCTEP problem, with the objective to minimise load curtailments \eqref{Objective: LC} or to minimise the total expected system cost \eqref{Objective: total cost}. The optimal solution is then analysed to perform a screening of potentially optimal investment options. That is, investments in line reinforcement and flexibility that contribute to the objectives \eqref{Objective: LC} or \eqref{Objective: total cost} are pre-selected for coalitional analysis. Note that if the optimal SCTEP solution identifies many investment options, e.g., more than 10 options, users are encouraged to select a limited number of investments to keep the coalitional analysis tractable.

In the second block, a cooperative game among the pre-selected investment options is formulated. That is, the set of players (investment options) and coalitions (combinations of investments) are defined. Then, to estimate the value of each coalition $S$, the SCTEP model \eqref{ACSCOPF: p_nm}-\eqref{ACSCOPF: curt limit}, \eqref{Objective: LC}, \eqref{Objective: total cost} is solved iteratively. At each iteration, the SCTEP model considers the set of available investment options corresponding to the set of players in coalition $S$, while all other investments are considered not available. Through this iterative process, the individual and combinatorial values of investments for the SCTEP problem are quantified.
Note that the cooperative game formulation does not introduce any additional constraints or binary variables into the SCTEP model. Thus, the optimisation problem for each coalition remains a continuous NLP.
Once the value of each coalition in $2^N$ is found, the marginal contributions of players to coalitions are calculated as defined in \eqref{CGT: MC_i}, and the Shapley value is computed using formula \eqref{CGT: Shapley}.

The last block of the framework includes the analysis of the calculated marginal contributions and the Shapley value for each player. This analysis ultimately allows to identify investments with the highest synergistic capabilities, i.e., the largest contributions in combination with other investments, and interpret the value of flexibility for SCTEP.

\section{Results and Discussion}\label{Section: results}
This section demonstrates the proposed flexibility valuation approach for two test transmission systems: a 5-bus illustrative network and a 30-bus UK system. For each system, the impacts of investments in line reinforcement and flexibility are analysed in terms of avoided load curtailment and expected system cost reduction. Then, insights into the value of flexibility, coalitional game formulations, and other modelling features are discussed. All simulations have been performed with JuMP 1.11.0 for Julia 1.6.1 language and Ipopt 3.14.4 solver.
Note that Ipopt, as well as other interior-point methods, does not guarantee global optimality for the continuous nonlinear problem \eqref{ACSCOPF: p_nm}-\eqref{ACSCOPF: curt limit}, \eqref{Objective: LC}, \eqref{Objective: total cost}. Nevertheless, it has been empirically demonstrated that Ipopt provides accurate and tractable solutions for realistic AC SCOPF case studies \cite{ALIZADEH2022Envisioning}.

\subsection{Case Study: Illustrative 5-bus System}
\begin{figure}
    \centering
    \includegraphics[width=0.9\columnwidth]{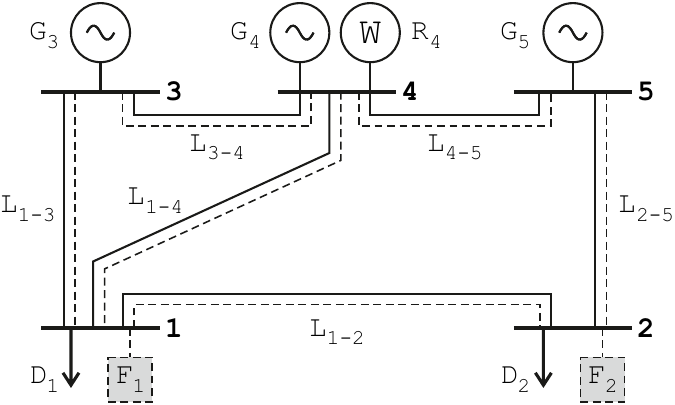}
    \caption{Case study: illustrative 5-bus system.}
    \label{Fig: C5 scheme}
\end{figure}

To illustrate the principles of flexibility valuation in SCTEP, the developed tool is first applied to a simple 5-bus system with 6 lines, as shown by its single-line diagram in Fig.~\ref{Fig: C5 scheme}. This system has been originally introduced in \cite{Florin_in_book2, ALIZADEH2022Envisioning} to test AC SCOPF models. There are three generators, each with a maximum capacity of 1500 MW and 750 MVar, supplying two loads (demand of 1100 MW and 400 MVAr at bus 1 and demand of 500 MW and 200 MVAr at bus 2).

A set of 6 contingencies is considered, corresponding to the tripping of every line (N-1 conditions). The 5-bus system is modified by setting a maximum transmission capacity of 800 MVA for each line, which makes some contingencies binding, i.e., resulting in load curtailment. To avoid potential load curtailment due to contingencies, it is necessary to upgrade the system and solve the SCTEP problem. It is assumed that the system planner has the following investment options: 1) each line can be reinforced by a maximum additional capacity of 100 MVA, as shown by the dashed lines in Fig.~\ref{Fig: C5 scheme}, 2) flexibility (e.g., energy storage systems) can be built at bus 1 and bus 2, with a maximum capacity of 100 MW.\footnote{It is assumed that large commercial flexible resources, e.g., battery energy storage systems, can be built and connected to the transmission network to support the system. Yet, in real cases, flexibility can be connected to distribution networks or even be located behind the meter \cite{Petrou2021}. Future research will include more detailed models to represent such flexible resources.}
The investment cost for flexibility, $c_{f}^\text{inv}$, and levelised cost of transmission lines, $c_{bm}^\text{inv}$, are assumed to be 5 \euro/MWh. Generators have quadratic cost functions as given in \cite{Florin_in_book2}, with the generator at but 3 being the cheapest one. Flexibility providers are expected to provide power at the price of 30 \euro/MWh. Penalty cost for load curtailment, $c_n^{\text{curt}}$, is set at $10^4$ \euro/MWh.

To introduce uncertainties in the system operation, bus 4 is assumed to have RES, a wind farm (denoted by ``W") with a maximum capacity of 500 MW and a generation profile given by two scenarios. The optimal solution of the SCTEP model \eqref{ACSCOPF: p_nm}-\eqref{ACSCOPF: curt limit} with cost-minimising objective function \eqref{Objective: total cost} recommends upgrading lines, 1-3, 1-4, 2-5, and investing in both flexibility providers to their maximum capacities. These investments allow reducing the total expected system cost by 8.5\%, from 0.621 to 0.568 mln\euro/h. However, this solution is not sufficient to interpret the value of flexibility in the considered planning problem. 
To deal with planning uncertainties, the system planner needs to prioritise the investment options and address the following questions before accepting the optimal expansion plan. How do flexibility providers contribute to avoiding potential load curtailment and reducing the total expected system cost? How effective are flexibility investments compared to line reinforcement? Which investment options jointly contribute to the defined objectives and thus have the highest synergistic capabilities in SCTEP? To address these questions and provide additional information for the system planner, the developed flexibility valuation tool was applied and the coalition analysis of the investment options was performed.

\begin{figure}
    \centering
    \includegraphics[width=\columnwidth]{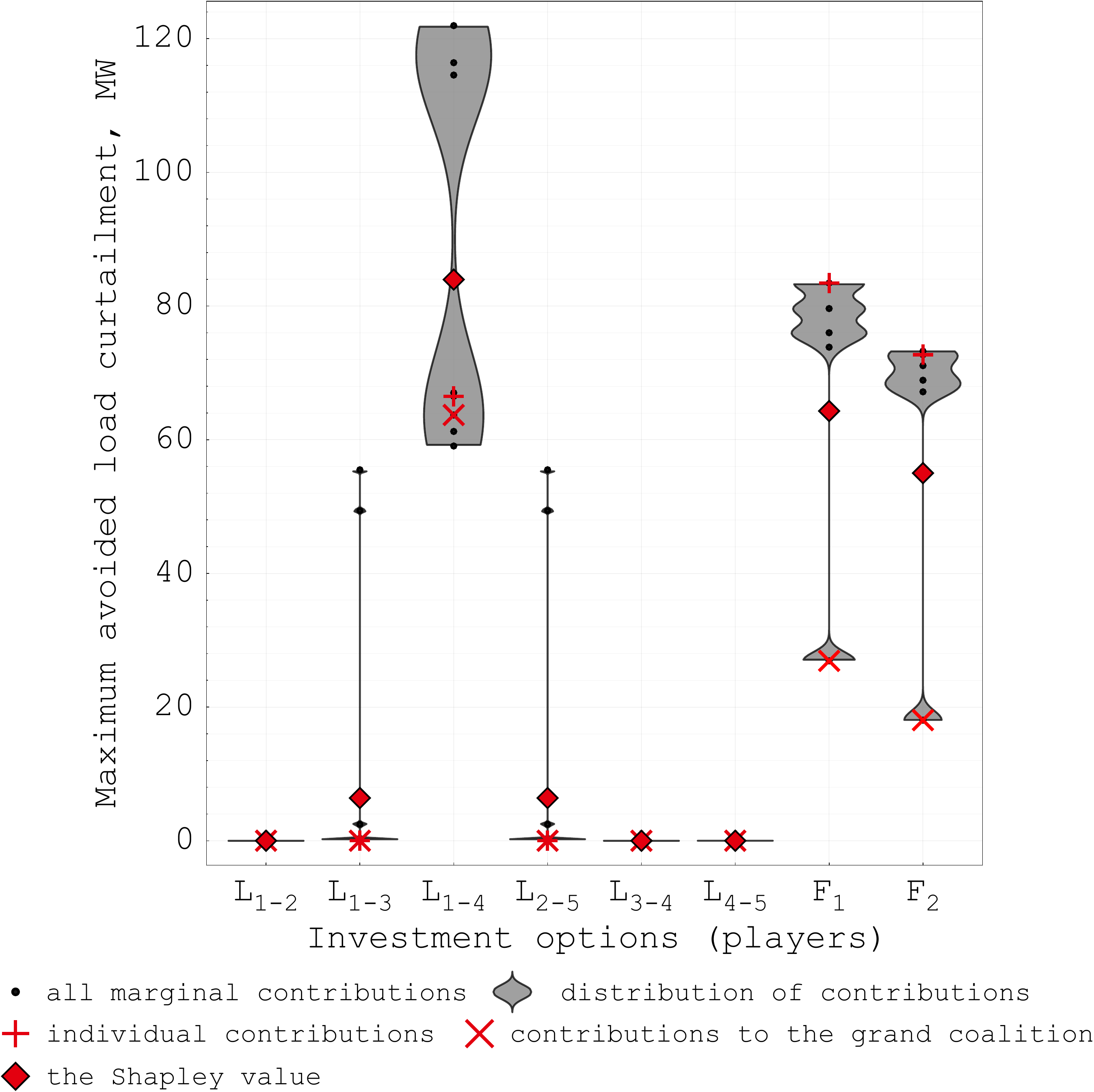}
    \caption{Valuation of flexibility in terms of the maximum avoided load curtailments for the 5-bus system. 
    }
    \label{Fig: C5 violin 1}
\end{figure}

The assumed investment options lead to a cooperative game with 8 players (six line reinforcements and two flexibility providers), which is feasible to analyse by considering a total of $2^8=256$ coalitions. First, the curtailment-minimising objective function \eqref{Objective: LC} was selected to solve the SCTEP model \eqref{ACSCOPF: p_nm}-\eqref{ACSCOPF: curt limit}. For each coalition, the maximum avoided load curtailment was estimated to characterise its value. Then, the marginal contributions of players were calculated using \eqref{CGT: MC_i} and the Shapley value \eqref{CGT: Shapley} was computed to represent the weighted average contribution to all possible coalitions. The results are presented in Fig.~\ref{Fig: C5 violin 1} as violin plots showing the distribution of the players' marginal contributions. As a single-valued estimation of contributions, the Shapley value is displayed by a single point (red diamond marker) for each player.


\begin{figure}[t]
    \centering
    \includegraphics[width=\columnwidth]{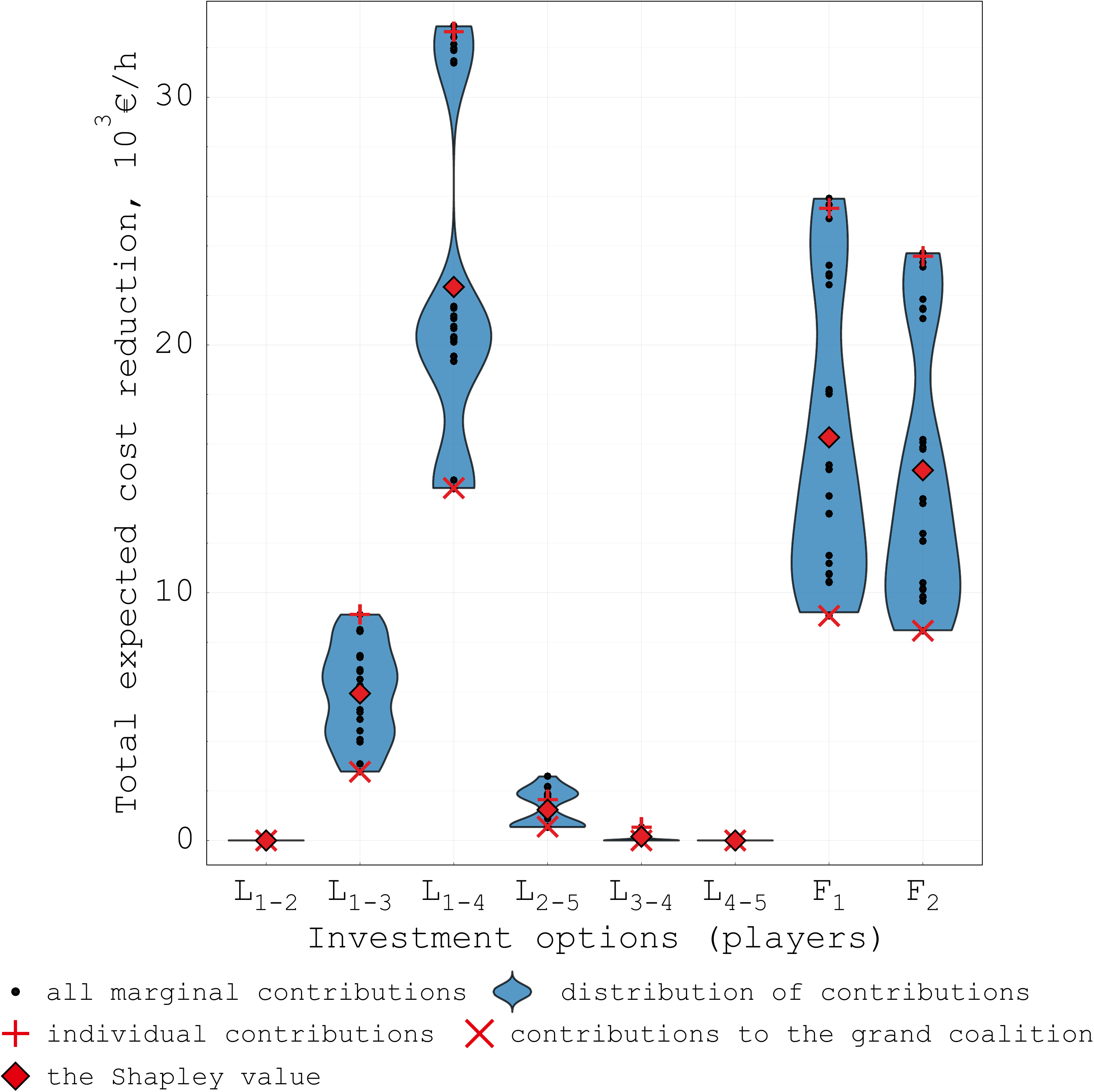}
    \caption{Valuation of flexibility in terms of the total expected cost reductions for the 5-bus system. 
    }
    \label{Fig: C5 violin 2}
\end{figure}

The performed coalition analysis demonstrates vast differences in the contributions of investment options to the maximum load curtailment. Reinforcements of lines 1-2, 3-4, and 4-5 make no positive contributions, regardless if considered individually or in combination with other investments. Therefore, they can be excluded from the investment portfolio. Lines 1-3 and 2-5 can contribute to avoiding load curtailment, but most of their contributions are low compared to other investment options. Line 1-4 appears to be the best investment option for load curtailment reduction, with the highest synergistic capability, that is, the largest contributions in combination with other investments.\footnote{Note that the maximum marginal contribution of line 1-4 exceeds its capacity of 100 MVA. This indicates that this line greatly contributes to other investment options which cannot maximise their curtailment reduction without line 1-4 reinforcement.} Flexibility providers at buses 1 and 2 can lead to significant load curtailment reductions and have a similar distribution of their contributions. It can be concluded that, if targeting the maximum load curtailment reduction, line 1-4 should be given priority in the system expansion planning. This justifies the need for a thorough coalitional analysis to prioritise investment options.

Second, objective function \eqref{Objective: total cost} was selected to solve the SCTEP model \eqref{ACSCOPF: p_nm}-\eqref{ACSCOPF: curt limit} while minimising the total expected system cost. Analysis of the players' contributions to the expected cost reductions is shown in Fig.~\ref{Fig: C5 violin 2}. Similar to the load curtailment analysis presented in Fig.~\ref{Fig: C5 violin 1}, line 1-4 and flexible units at buses 1 and 2 make the highest contributions, therefore offering the highest synergistic potential. Yet, the distribution of contributions is different as the cost-minimising simulations depend on cost assumptions and minimise both load and wind curtailment, as well as generation and investment costs. For example, line 1-3 makes consistent contributions to the total expected cost reduction by increasing the energy export from the cheap generator at bus 3.


The results illustrate how the value of flexibility in SCTEP can be interpreted as contributions to avoided load curtailment or total expected system cost reduction. Depending on the objective function and cost assumptions, the system planner can prioritise investments with higher synergistic capabilities and make well-informed decisions.

An important aspect of the performed simulations is the analysis of marginal contributions to all possible coalitions. Specifically, the range of all possible contributions is compared with the individual contributions of investments and their contributions to the grand coalition. This demonstrates that a simple sensitivity analysis, which excludes one of the investment options (or all but one investment), can lead to an incorrect assessment of the investment's value and its synergistic capability. 
For example, when analysing only contributions in the grand coalition and individual contributions to avoided load curtailment, the system planner may decide that flexibility at bus 1 should be prioritised as it has the highest contributions in the range [26.1, 83.3] MW. In comparison, line 1-4 has a range of contributions [63.7, 66.5] MW.
Yet, the entire range of contributions presented in Fig.~\ref{Fig: C5 violin 1} and the Shapley value reveal that reinforcement of line 1-4 is the most effective investment with the highest potential contributions and synergistic capability. Specifically, the maximum contribution to avoided load curtailment for line 1-4 is 121.9 MW and the Shapley value is 83.9 MW, while flexibility at bus 1 has the maximum contribution of 83.3 MW and the Shapley value of 64.3 MW.
Therefore, it is necessary to perform a thorough coalitional analysis to accurately prioritise investment options.

\subsection{Case Study: UK Electricity Transmission System}
\begin{figure}
    \centering
    \includegraphics[width=0.65\columnwidth]{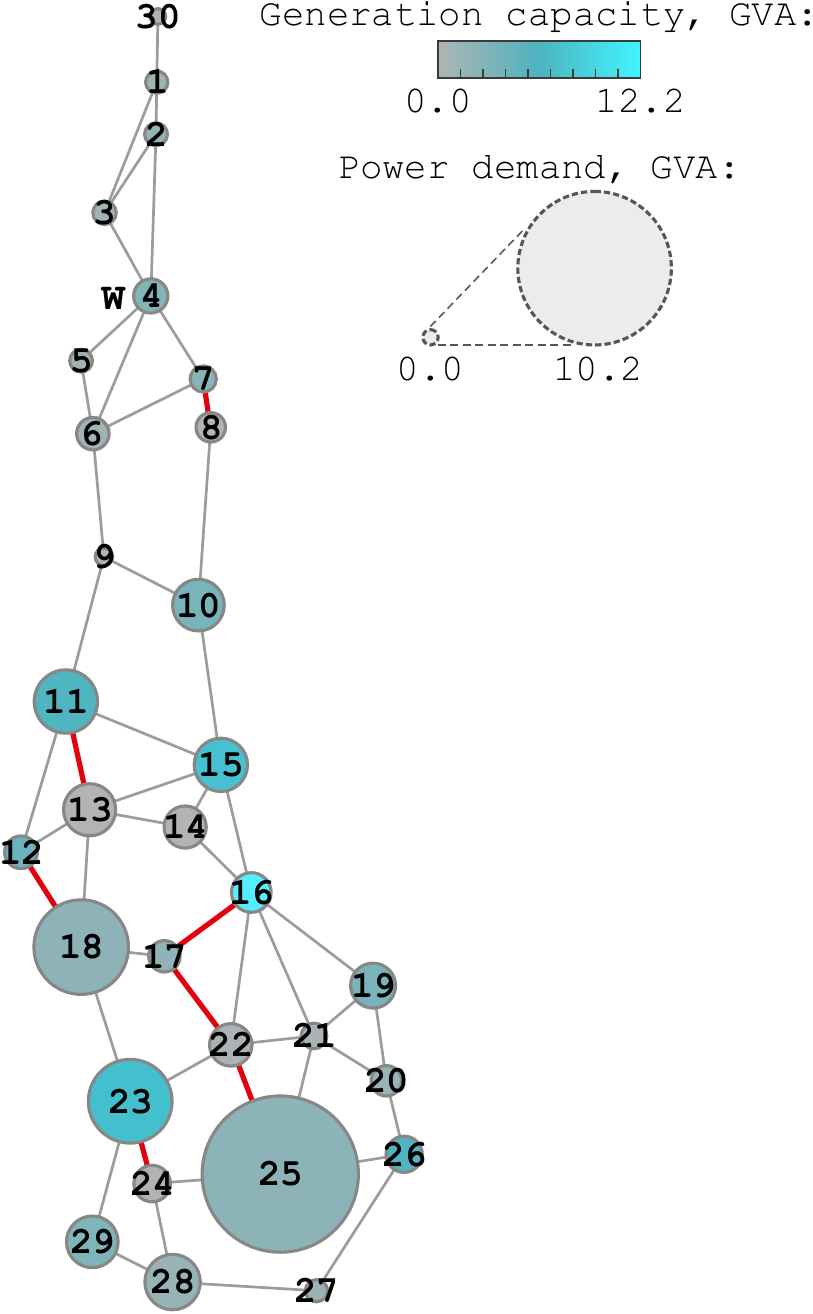}
    \caption{Case study: simplified electricity transmission system of the UK in 2020. Red lines indicate contingencies that become binding in 2050.}
    \label{Fig: UK scheme}
\end{figure}

To demonstrate the scalability of the flexibility valuation to larger systems, a simplified transmission system of the UK was selected \cite{martinez_cesena_UK_Tx}. This system, representing the UK transmission grid in 2020, has 30 buses, 100 lines, and a total power demand of 57.2 GW and 16.3 GVAr.\footnote{Note that the UK system has many parallel lines connecting the same buses. In the SCTEP, only one of the parallel circuits will be considered tripped to simulate contingencies and analyse N-1 security of the system.} The system is visualised in Fig.~\ref{Fig: UK scheme} as a graph where the size of nodes represents the nodal demand and the colour scheme indicates the location of generation capacity.
A set of 50 line contingencies (N-1 criterion) is considered in the SCTEP problem. To analyse a case with potential load curtailment due to contingencies, it is assumed that the planning horizon is 2050, where the total electricity demand will increase by 93\% compared to 2020 levels \cite{FES}. The total generation capacity is assumed to increase by 54\%, which is aligned with the UK's generation expansion plan in 2035. Thus, the planning problem corresponds to the planning between 2035 and 2050, when investments in line reinforcement and flexibility are needed to meet the growing power demand and guarantee N-1 security of the system.

Uncertainties associated with RES operation are introduced by considering 3 GW of wind capacity at bus 4 (denoted by ``W") given by a set of 10 scenarios.\footnote{The wind capacity assumption does not correspond to the total forecast wind generation in the UK, which is expected to reach dozens of GW. This assumption allows to demonstrate explicit modelling of uncertain generation in SCTEP, focusing on load curtailment rather than wind curtailment.}
Similar to the 5-bus system case study, the investment cost for flexibility, $c_{f}^\text{inv}$, and levelised cost of transmission lines, $c_{bm}^\text{inv}$, are assumed to be 5 \euro/MWh. Generators have linear cost functions in the range of 1-70 \euro/MWh. Flexible power price is 30 \euro/MWh. Penalty cost for load curtailment, $c_n^{\text{curt}}$, is set at $10^4$ \euro/MWh.

In total, 51 lines are assumed available for reinforcement by a maximum additional capacity of 1000 MVA, and all buses are considered for flexibility investments with a maximum capacity of 1000 MW. This leads to a set of 81 investment options, which are too many for a coalitional analysis. Therefore, a screening of the initial investment options was performed by measuring the individual contributions of these options to maximum avoided load curtailment. That is, SCTEP model \eqref{ACSCOPF: p_nm}-\eqref{ACSCOPF: curt limit}, \eqref{Objective: LC} was solved for coalitions of one player.
Among the line reinforcement options, only three lines, 7-8, 11-13, and 17-16, have positive contributions and therefore are selected for further analysis. All 30 flexibility locations have positive impacts on avoided load curtailment. Among them, the four locations with the highest marginal contributions, buses 25, 26, 27, and 28, are selected. The resulting cooperative game has 7 players and can be solved by analysing $2^7=128$ coalitions.

\begin{figure}
    \centering
    \includegraphics[width=\columnwidth]{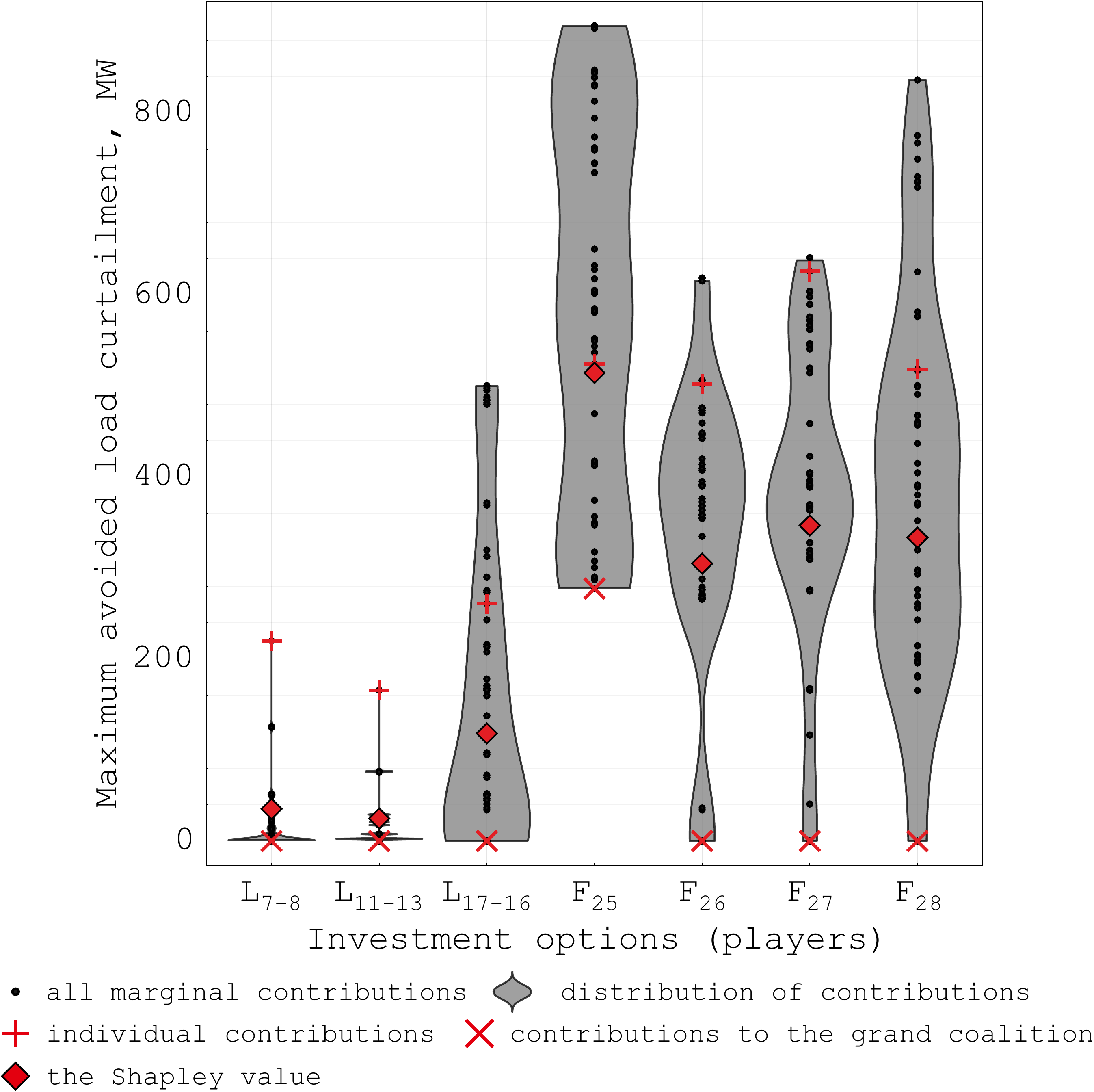}
    \caption{Valuation of flexibility in terms of the maximum avoided load curtailments for the UK transmission system.
    }
    \label{Fig: UK violin 1}
\end{figure}

Analysis of the marginal contributions to avoided load curtailment for the selected investment options is presented in Fig.~\ref{Fig: UK violin 1}. It can be concluded that investing in the flexibility provider at bus 25 should be prioritised as this option has the highest synergistic capability to avoid load curtailment. Among the line reinforcements, line 17-16 should be prioritised as it makes consistent positive contributions.

\begin{figure}[t]
    \centering
    \includegraphics[width=\columnwidth]{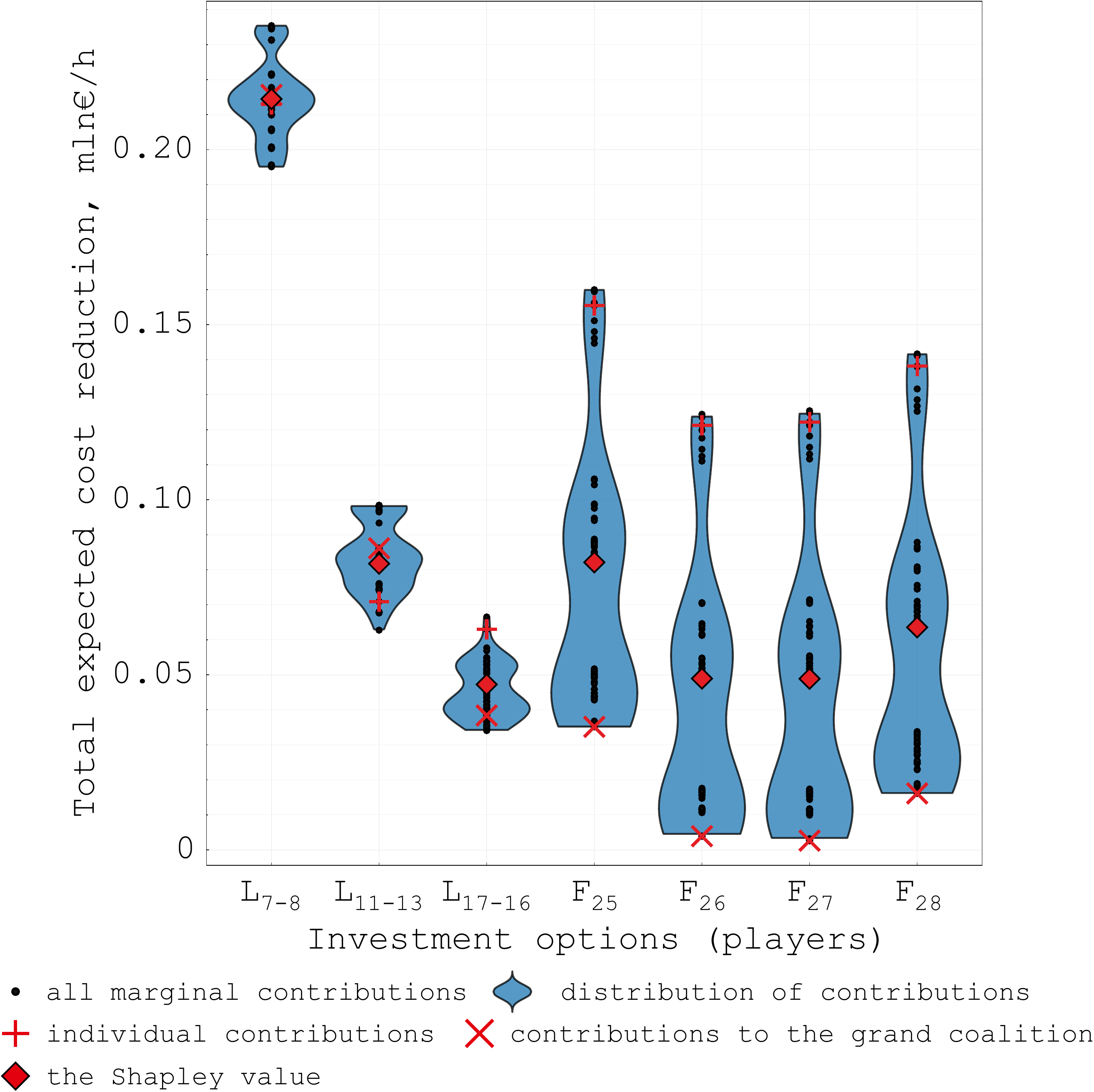}
    \caption{Valuation of flexibility in terms of the total expected cost reductions for the UK transmission system.
    }
    \label{Fig: UK violin 2}
\end{figure}

The cost-minimising SCTEP model \eqref{ACSCOPF: p_nm}-\eqref{ACSCOPF: curt limit}, \eqref{Objective: total cost} recommends investments in 12 line reinforcements and 8 flexibility providers, with the optimal capacities ranging between 22-1000 MW. The optimal expansion plan allows to reduce the total expected cost by 3\%, from 28.6 to 27.7 mln\euro/h. Analysis of the marginal contributions to the total expected system cost reduction for the selected investment options is presented in Fig.~\ref{Fig: UK violin 2}. In contrast to the load curtailment analysis, line reinforcement options make more significant contributions in terms of costs. For example, line 7-8 has the highest contributions to the total expected cost reduction since it reduces wind curtailment at bus 4 and enables the transfer of cheaper power from the northern part of the system. It follows that, if focusing on the total expected cost minimisation, the system planner should prioritise reinforcing this line.


\subsection{Discussion}
The developed SCTEP tool enables interpreting the value of flexibility as contributions to avoided load curtailment or total expected system cost reduction and comparing it against line reinforcements.
This information allows system planners to prioritise investments with higher contributions and synergistic capabilities.
The proposed flexibility valuation approach is not only a post-processing of the optimal SCTEP solution, but a comprehensive analysis of multiple solutions corresponding to various coalitions of investment options. Solving security-constrained problems is known to be computationally expensive for large systems with many scenarios \cite{ALIZADEH2022Envisioning}. Altough computation time is not a concern at the planning stage, the scalability of the proposed approach is a major limitation, as $2^N$ coalitions have to be simulated to estimate all contributions for the selected investment options.

Nevertheless, there exist two ways to overcome the scalability issue and apply the developed SCTEP tool to large systems. First, as simulations of different coalitions are not interdependent, the process can be greatly parallelised to compute the value of coalitions using multiple cores and processors, thus reducing the computational time for coalition analysis.
Second, recent studies on the Shapley value applications have demonstrated that coalitional analysis can be performed accurately enough by considering a strategically selected limited number of coalitions \cite{Jia2020,Mitchell2022,Cremers2023}. That is, a truncated coalitional structure can be used to estimate players' contributions and approximate the Shapley value.
Moreover, coalitional analysis can be performed in several stages, where investment options are screened by limiting the number of coalitions. Then, the options with less significant contributions are discarded, and the analysis is repeated for a cooperative game with fewer players.

\section{Conclusion}\label{Section: conclusion}
Accurate N-1 secure planning of future transmission systems is becoming increasingly important as the integration of uncertain RES accelerates. To guarantee reliable and cost-effective system operation, system planners exploit SCTEP models and optimise investments in line reinforcement and flexibility.
Being a complex optimisation problem, SCTEP is typically solved once to find the optimal investment strategy. However, a single optimal solution is not sufficient to interpret the value of flexibility in transmission planning and support well-informed decision making. 

In this regard, this work introduces a SCTEP tool that enables interpreting the value of flexibility in terms of avoided load curtailment and total expected system cost reduction.
Inspired by cooperative game theory, the tool ranks the contributions of flexibility providers in SCTEP and compares them against traditional line reinforcements.
This information can be used by system planners to prioritise investments with higher contributions and synergistic capabilities.
The proposed flexibility valuation approach is demonstrated for two transmission systems (illustrative 5-bus system and UK's 30-bus system).
It is found that, depending on the objectives considered by the system planner, investments in flexibility and line reinforcement can have vastly different values. Flexibility providers tend to make consistent contributions to avoided load curtailment, while line reinforcement can outperform investments in flexibility in terms of total expected system cost reduction.
The distribution analysis of the players' contributions also showed that individual contributions of investment options cannot always be used to correctly estimate the entire range of possible contributions.
That is, a simple sensitivity analysis for traditional planning models (exclusion of certain investments) can lead to an incorrect assessment of the investment's value and its synergistic capability. This justifies the need for a thorough coalitional analysis to prioritise investment options.

The aim of this work is to illustrate the concept of flexibility valuation in SCTEP and explain the usefulness of this information for system planning. Therefore, not all the computational aspects have been thoroughly analysed. Future work will further investigate the computational aspects and scalability of the proposed approach by considering the Shapley value approximations and testing larger systems with multiple uncertainties, including multi-period optimisation with intertemporal constraints and energy storage systems. 

\section*{Acknowledgements}
The authors would like to thank (i) the European Union’s
Horizon 2020 research and innovation programme for their financial support under grant agreement No. 864298 (project ATTEST) and (ii) the UK's EPSRC for their financial support under grand agreement EP/W019795/1 (project ModFlex).



%




\bibliography{references.bib}
\bibliographystyle{IEEEtran}


\end{document}